\documentclass[11pt,twoside]{article}
\usepackage{asp2010}

\resetcounters

\begin{document}

\title{Towards a standardised line list for G191-B2B, and other DA type objects.}
\author{S. P. Preval,$^1$ M. A. Barstow,$^1$ J. B. Holberg,$^{1,2}$ N. J. Dickinson$^1$
\affil{$^1$Physics Building, University of Leicester, University Road, Leicester, LE1 7RH, UK}
\affil{$^2$Lunar and Planetary Laboratory, Gould-Simpson Building, University of Arizona, Tucson, AZ 85721, USA}}

\begin{abstract}
We present a comprehensive analysis of the far UV spectrum of G191-B2B over the range of 900-1700\AA\, using co-added data from the FUSE and STIS archives. While previous identifications made by \cite{holberg03a} are reaffirmed in this work, it is found that many previously unidentified lines can now be attributed to Fe, Ni, and a few lighter metals. Future work includes extending this detailed analysis to a wider range of DA objects, in the expectation that a more complete analysis of their atmospheres can be realised.
\end{abstract}

\section{Introduction}
G191-B2B is one of the best studied objects of it's kind. However, while we have extensive knowledge of the composition of the white dwarf's atmosphere, some aspects remain unknown. For example, the majority of the absorption features present in the object's UV spectrum have been identified, while some features remain elusive. For white dwarfs whose effective temperature $T_\mathrm{eff}$ exceeds 50,000K, determination of $T_\mathrm{eff}$ becomes ambiguous due to the lyman-balmer line problem \citep{barstow03a}, where determinations of $T_\mathrm{eff}$ using lyman or balmer line profiles can differ by 1000's of K. A similar effect is not seen in determining the surface gravity $\textrm{log }g$. The lyman-balmer line problem may be due to incompleteness in the models used in calculating $T_\mathrm{eff}$ and $\textrm{log }g$, and it is suggested in this paper that a solution may be to include additional opacities in model atmosphere  calculations.

\section{New high quality data}
\subsection{High signal to noise spectra}
Using the Mikulski Archive for Space Telescopes\footnote{http://archive.stsci.edu} (MAST), \cite{barstow10a} co-added all available low resolution (LWRS) FUSE exposures, and \cite{holberg03a} co-added all of the high resolution eschelle STIS spectra to cover the entire NUV and FUV range. Wavelength ranges, number of exposures, and gratings are summarised in Table \ref{table:observe}. The resulting spectra had exceptionally high signal to noise ($\approx{200}$ at some wavelengths), allowing features with equivalent widths as small as 2m\AA\, to be detected.

\begin{table}
\centering
\begin{tabular}[H]{@{}cccc}
\hline
Instrument: & & No of & Data reduction \\
Filter & Range (\AA) & exposures & reference\\
\hline
FUSE: & & &  \\ 
LWRS & 910 - 1185 & 57 & \cite{barstow10a} \\
STIS: & & & \\
E140H & 1150 - 1700 & 22 & \cite{holberg03a} \\
STIS: & & & \\
E230H & 1600 - 3165 & 40 & \cite{holberg03a} \\
\hline
\end{tabular}
\caption{A summary of the data used in the co-addition process. The references provided detail how the data was reduced.}
\label{table:observe}
\end{table}

\subsection{New atomic data}
In this analysis, two line lists were compiled using different data releases from the Kurucz and Kentucky atomic databases. The first line list makes use of the 2006 data release from Kurucz \citep{kurucz06a}, while the second line list combines the 2011 and 2012 data releases from the Kurucz\footnote{http://kurucz.harvard.edu/} and Kentucky\footnote{http://www.pa.uky.edu/~peter/newpage/} atomic databases respectively. Where entries of the Kurucz and Kentucky databases conflicted, the Kurucz entry was favoured owing to the greater accuracy of the oscillator strengths ($f$-value) provided. Lines that did not have a corresponding $f$-value in the Kentucky database were nominally set at $1.00\times{10^{-6}}$. The combined line list increases the number of transitions considered in the synthetic spectrum calculation, while also improving on the accuracy of transitions in previous data releases.
\begin{figure}[t]
\centering
\includegraphics[clip=true,trim = 200sp 360sp 1187sp 714sp, width=120mm]{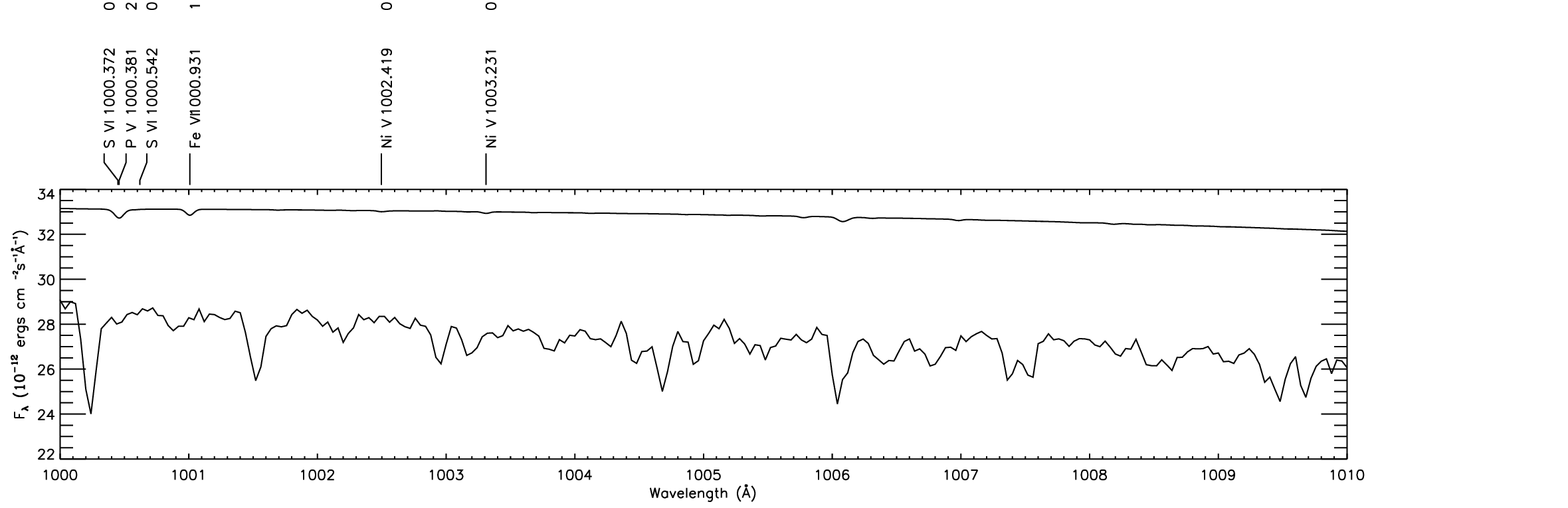}
\includegraphics[clip=true,trim = 200sp 360sp 1187sp 714sp, width=120mm]{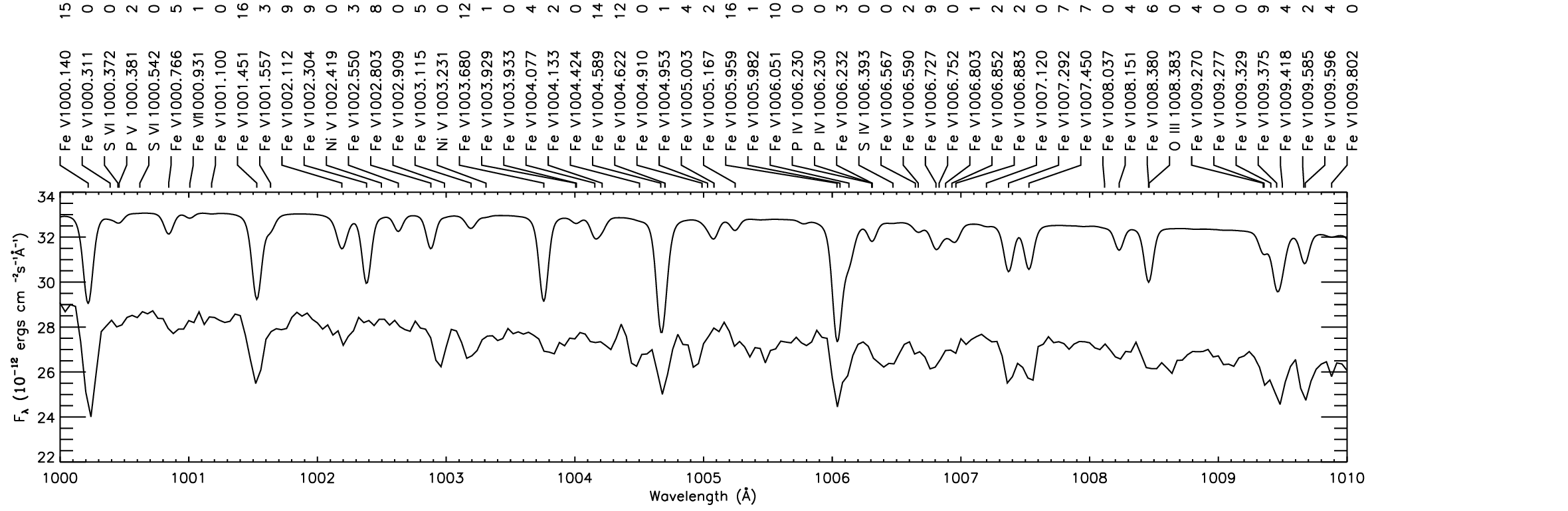}
\caption{A comparison between the 2006 Kurucz and the 2012 combined Kentucky + Kurucz 2011 datasets. The bottom and top lines in each plot are the observed and synthetic spectra respectively, where the latter has been offset for clarity.}
\label{fig:newlist}
\end{figure}

\begin{figure}[t]
\centering
\includegraphics[clip=true,trim = 54sp 360sp 735sp 690sp, width=100mm]{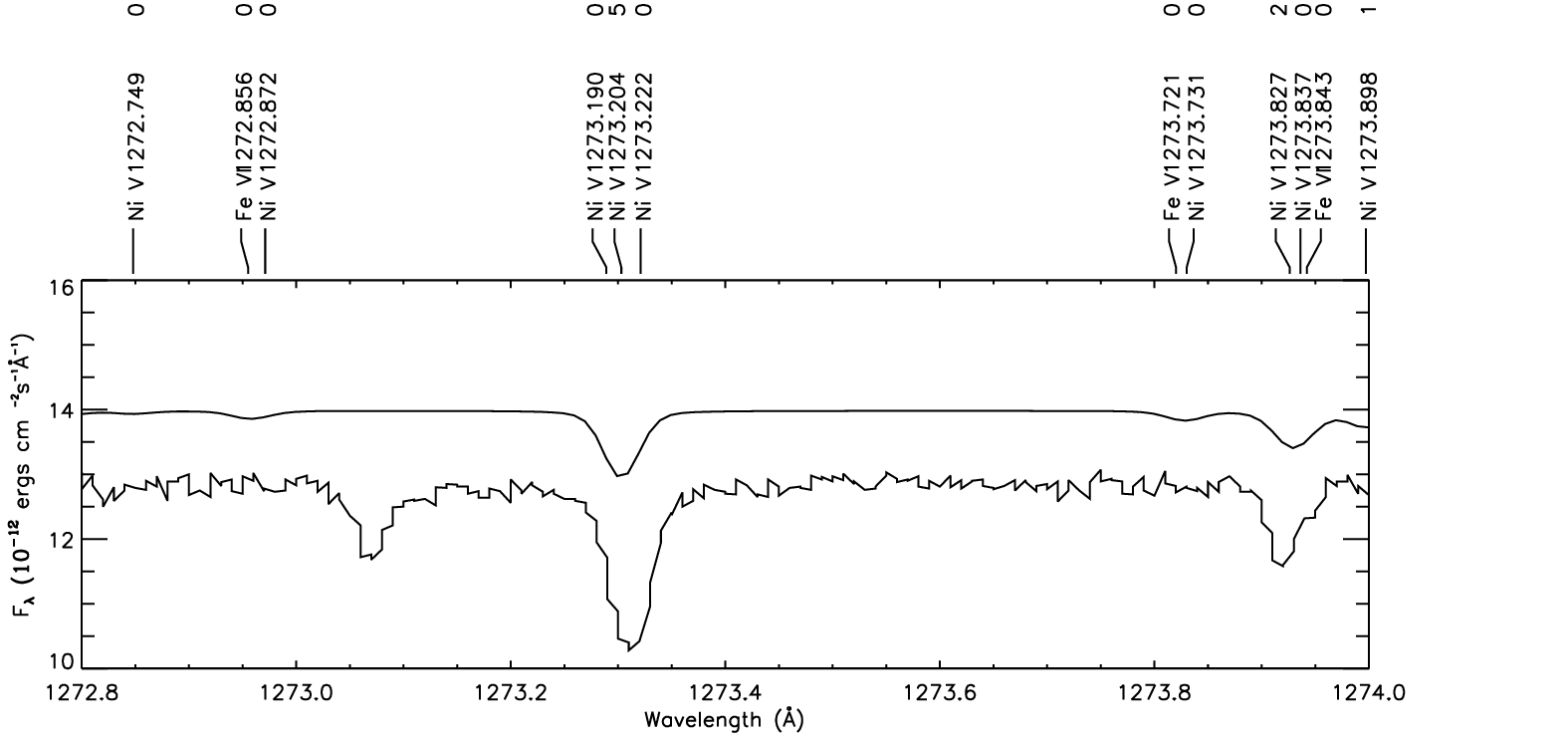}
\includegraphics[clip=true,trim = 54sp 360sp 735sp 690sp, width=100mm]{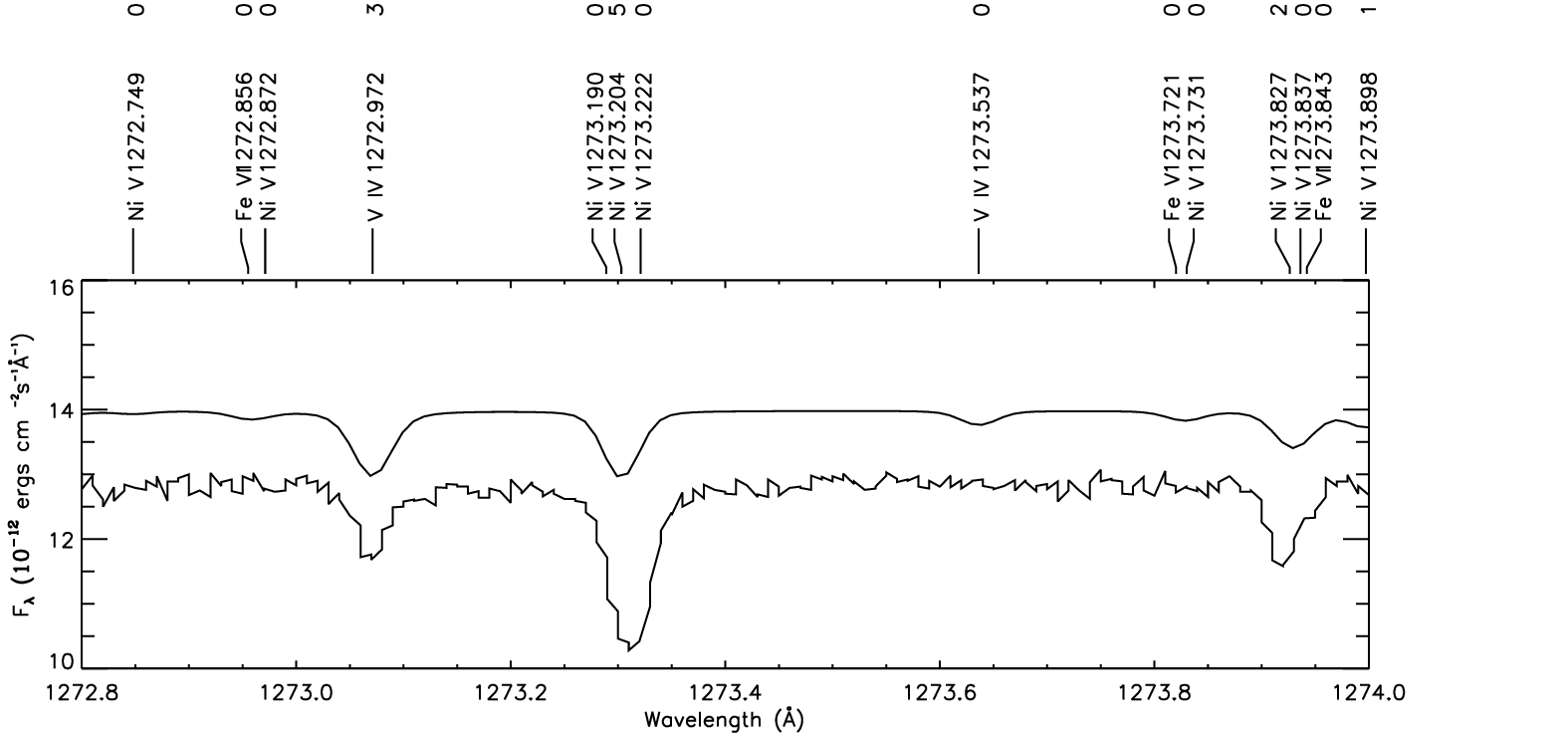}
\caption{An unidentified line at $\lambda\lambda{1273.069}$, a potential identification being V IV. The top plot shows a non detection when a V abundance of $1.07\times{10^{-9}}$ is assumed. The bottom plot shows the V line when an abundance of $1.00\times{10^{-6}}$ is assumed. The bottom and top lines in each plot are the observed and synthetic spectra respectively, where the latter has been offset for clarity.}
\label{fig:vanad}
\end{figure}

\section{Modelling}
743 absorption features were measured in the near and far UV spectrum of G191-B2B to greater than $5\sigma$. Line identification has been performed by comparing synthetic spectrum calculations to observed spectral flux distributions from G191-B2B. A model atmosphere using abundances from \cite{barstow03a} was calculated by TLUSTY \citep{hubeny88a}, and synthesised with SYNSPEC \citep{hubeny11a} using the two line lists described in \textsection{2.2} from different data releases. A comparison of the synthesised spectra using the two line lists is depicted in Figure \ref{fig:newlist}, where the top and bottom plots compare the synthetic and observed spectra using the old and new line lists respectively (all plots were produced using an IDL package called SYNPLOT\footnote{http://nova.astro.umd.edu/Synspec43/synspec.html}. In the top plot of Figure \ref{fig:newlist}, there are few similarities between the predicted and observed line profiles, whereas the bottom plot depicts the tentative identifications of several Fe lines among others. The line positions appear to correspond quite accurately with observation, however, the line depths for the new list occasionally descend deeper than the observed absorption profiles. This disagreement is likely to be due to the f-values (described in \textsection{2.2}) assigned to transitions whose real $f$-value is unknown. Some absorption features however have remained unidentified. One example is a line measured at $\lambda\lambda{1273.069}$, depicted in Figure \ref{fig:vanad}. One possible identification originates from that of V IV $\lambda\lambda{1272.972}$. The top and bottom plot in Figure \ref{fig:vanad} depicts the synthetic spectrum with V at abundances of $1.07\times{10^{-9}}$ and $1.00\times{10^{-6}}$ relative to H respectively.

\section{Conclusion}
We have performed an in-depth survey of the near and far UV spectrum of G191-B2B. We have measured 743 absorption features to $5\sigma$ confidence, whose identifications will be detailed in future work. Utilising a new line list combining the 2011 and 2012 data releases from Kurucz and Kentucky, we have identified the majority of these lines to originate from Fe IV, V, and VI, with occasional detections of lighter metals. Using the newly identified lines, it is now possible to calculate revised abundances of Fe and Ni in future model atmosphere calculations. The obvious increase in identified Fe absorption features presents the question of opacity consideration in model atmosphere calculations. With the resulting changes in the atmospheric structure from the inclusion of additional opacities, it may also be possible to explain the discrepancies between the measured values of $T_\mathrm{eff}$ obtained from Lyman and Balmer lines by \cite{barstow03a}.

\acknowledgements SPP, MAB, and NJD gratefully acknowledges the continued support of the Science and Technology Facilities Council (STFC). JBH acknowledges a visiting professorship from the University of Leicester.
\bibliography{PREVAL_POSTER}

\end{document}